\documentclass[11pt,a4paper]{article}

\usepackage[a4paper,left=3cm,right=3cm,top=2.5cm,bottom=3.5cm]{geometry}
\usepackage[utf8]{inputenc} 
\usepackage[T1]{fontenc}    
\usepackage{hyperref} 
\hypersetup{
    colorlinks=true,
    linkcolor=blue,
    citecolor=blue,
    filecolor=blue,
    urlcolor=blue}
\usepackage{authblk}

\usepackage{url}            
\usepackage{booktabs}       
\usepackage{amsfonts}       
\usepackage{nicefrac}       
\usepackage{microtype}      
\usepackage{graphicx}
\usepackage{xcolor}
\usepackage{framed}
\usepackage{tcolorbox}
\usepackage{orcidlink}
\tcbuselibrary{breakable}
\usepackage[
    backend=biber,
    sorting=none,
    style=numeric-comp,
    giveninits=true
]{biblatex}
\usepackage{doi}
\usepackage{microtype}

\definecolor{boxcolframe}{rgb}{0.78,0.78,0.78}
\definecolor{boxcolbg}{rgb}{0.9,0.9,0.9}
\newtcolorbox[auto counter]{pasqbox}[2][]{
colback=boxcolbg, colframe=boxcolframe, arc=0pt, leftrule=0pt, rightrule=0pt, bottomrule=0pt, toprule=0pt, titlerule=0pt, fonttitle=\bfseries, coltitle=black, toptitle=5pt, bottomtitle=5pt, title={Box \thetcbcounter $~|$ #2}, #1}

\usepackage{braket}
\usepackage{amssymb}
\usepackage{amsmath}
\usepackage{enumitem}

\usepackage{array}
\usepackage{arydshln}

\usepackage{hhline}

\newcommand{\Graph}{{\mathcal{G}}}

\title{\vspace{-1.3cm}\Large\textbf{Graph Algorithms with Neutral Atom Quantum Processors}}
\author[1]{Constantin Dalyac\thanks{These authors contributed equally to this work}\,\orcidlink{0000-0002-0339-6421}}
\author[1,2]{Lucas Leclerc$^*$\,\orcidlink{0000-0003-0581-9165}}
\author[1]{Louis$~$Vignoli\, \orcidlink{0000-0003-1806-9762}}
\author[1]{Mehdi$~$Djellabi}
\author[1]{Wesley$~$da$~$Silva$~$Coelho}
\author[1]{Bruno$~$Ximenez}
\author[1]{Alexandre$~$Dareau\,\orcidlink{0000-0001-7581-4701}}
\author[1]{Davide$~$Dreon\,\orcidlink{0000-0002-4831-272X}}
\author[1]{Vincent$~$E.$~$Elfving\,\orcidlink{0000-0002-5105-5664}}
\author[1]{Adrien$~$Signoles\,\orcidlink{0000-0001-7822-9444}}
\author[1]{Louis-Paul$~$Henry}
\author[1]{Loïc$~$Henriet\thanks{\{constantin,loic\}@pasqal.com}}

\affil[1]{PASQAL, 7 rue Léonard de Vinci, 91300 Massy, France}
\affil[2]{Laboratoire Charles Fabry, Institut d'Optique Graduate School, Universit\'e Paris-Saclay, CNRS, 91127$~$Palaiseau, France}

\date{}
\begin{document}
\maketitle
\vspace{-1cm}

\begin{abstract}
Neutral atom technology has steadily demonstrated significant theoretical and experimental advancements, positioning itself as a front-runner platform for running quantum algorithms. One unique advantage of this technology lies in the ability to reconfigure the geometry of the qubit register, from shot to shot. This unique feature makes possible the native embedding of graph-structured problems at the hardware level, with profound consequences for the resolution of complex optimization and machine learning tasks. By driving qubits, one can generate processed quantum states which retain graph complex properties. These states can then be leveraged to offer direct solutions to problems or as resources in hybrid quantum-classical schemes. In this paper, we review the advancements in quantum algorithms for graph problems running on neutral atom Quantum Processing Units (QPUs), and discuss recently introduced embedding and problem-solving techniques. In addition, we clarify ongoing advancements in hardware, with an emphasis on enhancing the scalability, controllability and computation repetition rate of neutral atom QPUs.
\end{abstract}

\section{Introduction}
The current landscape of quantum processors demonstrates significant strides forward in terms of computational capabilities \cite{Preskill18,Ezratty23}. Among the various available technologies, neutral atom quantum processors have emerged as particularly promising due to their potential to offer versatile qubit platforms\,\cite{Kim2023,Henriet2020}. These platforms offer a high level of control, allowing the successful tackling of quantum simulation problems that are challenging to emulate classically. Recent setups have achieved high-fidelity simulations, including quantum Ising models with up to 200 spins\,\cite{Browaeys2020_many_body} and quantum XY models with 100 spins in two-dimensional arrays of atoms\,\cite{Chen2023}. Nowadays the continuous enhancement of operation fidelities\,\cite{Scholl23, Evered23, Ma2023} signifies the ongoing progress in neutral atom quantum processors. As such, the unique attributes of quantum systems can be harnessed for specialized computations in groundbreaking approaches.

One such paradigm-shifting avenue lies in the utilization of neutral atom quantum processors to address complex graph problems, spanning from optimization challenges\,\cite{abbas2023quantum} to machine learning tasks\,\cite{Biamonte_2017}. Intrinsic to a multitude of scientific, engineering and industrial scenarios, graphs represent complex connections between entities. The practical importance of these concepts becomes apparent in various scenarios, such as social networks where individuals are connected based on their relationships, in logistics where transportation routes constitute edges connecting significant locations, or in molecular biology where edges represent chemical bonds between atoms. The abstractions that graphs provide make them invaluable tools for addressing and understanding complex problems. However, the extensive flexibility they offer can also lead to resource-intensive data analyses. Harnessing the innate properties of neutral atoms, particularly the phenomenon of Rydberg blockade\,\cite{Lukin_2001} and the native ability to perform interacting quantum walks\,\cite{scholl2022microwave}, offers an appealing avenue for achieving enhanced computational capabilities in graph-related endeavors. 

 As hardware improvements (detailed in boxes throughout the text) lead to higher numbers of qubits\,\cite{schymik2022situ, pause2023supercharged} and as new features such as partial addressability\,\cite{zhang2023scaled} gradually become available, more and more industrial graph-problems can be embedded in models accessible to neutral atom processors. By leveraging this enhanced controllability, neutral atom technology can help to swiftly and efficiently navigate complex computational challenges.

\section{Graph Hamiltonians with interacting neutral atoms in tweezer arrays}

Neutral atom in arrays of optical tweezers have been shown to be a robust platform for implementing quantum Hamiltonians and carrying out analog quantum processing. A distinctive strength of this technology lies in its capacity to program arbitrary register geometries in both 2D and 3D and to foster strong programmable interactions among qubits.
The adaptability of the register geometry arises from employing holographic techniques to design the arrangement of microscopic atomic traps\,\cite{Nogrette14,Lee16}, usually separated by few micrometers.
Furthermore, the atoms when promoted to Rydberg states can exhibit power-law two-body interactions derived from dipole-dipole couplings. By carefully tuning their respective pair distances, one can then program the magnitude of their interactions and precisely engineer an interaction Hamiltonian embedding the adjacency matrix of a graph $\Graph$.

\begin{figure}[ht]
\center
\includegraphics[width=\textwidth]{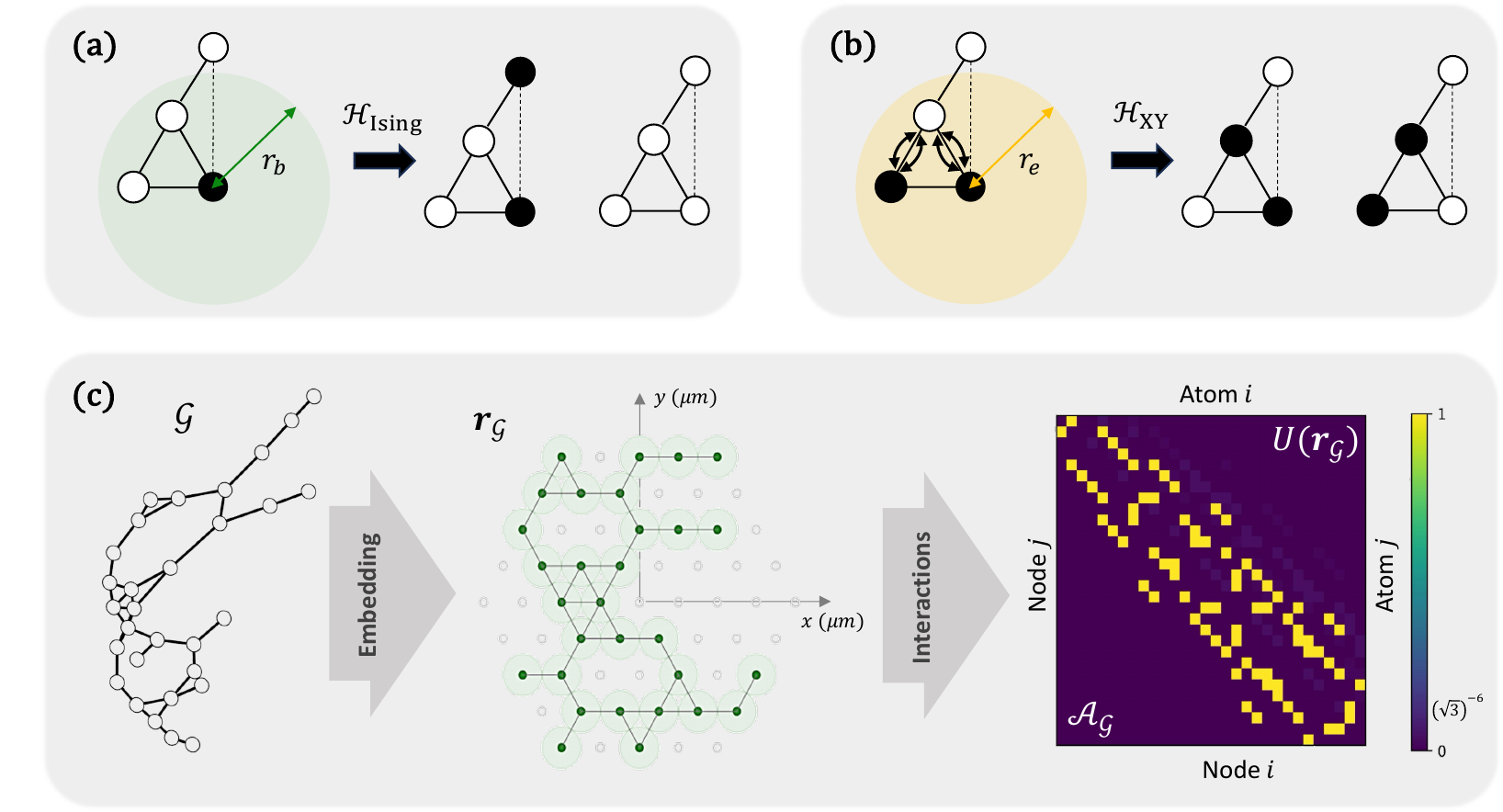}
\caption{ Depending on the atomic states involved, the qubits experience a different interaction Hamiltonian (Ising or XY). \textbf{(a)} Under resonant driving in an Ising Hamiltonian, an atom excited in a Rydberg state (black dot) prevents the excitation of all the atoms contained in a disc of radius $r_b$ for which the interaction is stronger than the excitation strength (full line). On the other hand, the interaction with an atom at a distant larger than $r_b$ is weak (dotted line). The evolution under the Ising Hamiltonian can therefore see the initial excitation removed or make the further atom gain another excitation.
\textbf{(b)} Under the spin-exchange term, an excitation can hop on neighboring vertices along a graph edge.
Their spin nature prevents two particles from being on the same site, therefore leading to an interacting version of a continuous quantum walk. \textbf{(c)}
In an adequate spatial layout of the atoms, the interactions can reproduce the adjacency matrix $\mathcal{A}_\Graph$ of an input graph $\Graph$. The sharp decay of the interaction allows neglecting terms between second (and further) nearest neighbors.}
\label{fig1}
\end{figure}

The interaction, that maps to spin Hamiltonian, can change depending on the Rydberg levels involved. One possibility is to choose the $\ket{0}$-state as the ground state of the electronic structure of the atom and the $\ket{1}$-state as a highly-excited Rydberg state. In this case, two atoms prepared in the $\ket{1}$-state interact through a van der Waals interaction scaling as $C_6 / r^6$, where $r$ corresponds to the interatomic distance and $C_6$ is a constant that depends on the chosen Rydberg level.
For a system of $N$ atoms the van der Waals interactions are pairwise additive and contribute to the Ising Hamiltonian
\begin{equation}
    \mathcal{H}_{\text{Ising}}(t)/\hbar =  \frac{ \Omega(t)}{2} \sum_{i=1}^N \sigma_i^x - \delta(t) \sum_{i=1}^N n_i + \sum_{i<j} \frac{C_6/\hbar}{r_{ij}^6}n_in_j,
\label{eq:ising_hamiltonian}
\end{equation}
with $\sigma^{\alpha}_j$ ($\alpha=x,y,z$) the Pauli matrices associated to qubit $j$ and $n_j=(\mathbb{I}+\sigma_j^z)/2$.
The first two terms are induced by a laser that couples the qubit states, where the Rabi frequency $\Omega(t)$ and detuning $\delta(t)$ are the parameters that control the transverse and longitudinal components respectively.
One can leverage the Ising interaction to engineer the well-known Rydberg blockade effect where two Rydberg atoms experience an interaction strong enough to shift significantly the energy of the doubly excited state, preventing their simultaneous excitation~\cite{jaksch_fast_2000,Urban2009,Gaetan2009}. Under resonant ($\delta=0$) drive with constant Rabi frequency $\Omega$, two atoms separated by a distance smaller than the Rydberg blockade radius $r_b$, defined as $r_b = (C_6/\hbar \Omega)^{1/6}$, cannot be both excited at the same time (Fig.\,\ref{fig1}a).
However, the rapid decay of the interaction enables us to neglect the interaction term for atoms located beyond the blockade radius (depicted as a dotted line in Fig. \ref{fig1}a).

The dipole-dipole interactions manifesting in Rydberg systems can also lead to an Hamiltonian featuring spin-exchanges. By encoding the qubit states $\ket{0}$ and $\ket{1}$ into two different dipole-coupled Rydberg states, the interaction Hamiltonian exhibit particle-exchange terms inducing a transformation from the pair state $\ket{0_i1_j}$ to $\ket{1_i0_j}$. For a system of $N$ atoms, these interactions contribute to the XY Hamiltonian
\begin{equation}
\mathcal{H}_{\text{XY}}(t)/\hbar = \frac{ \Omega(t)}{2} \sum_{i=1}^N \sigma_i^x - \delta(t) \sum_{i=1}^N n_i + \frac{1}{2}\sum_{i < j}\frac{C_3/\hbar}{r_{ij}^3} \left(\sigma_i^x \sigma_j^x + \sigma_i^y \sigma_j^y \right).
\label{eq:xy_hamiltonian}
\end{equation}
The exchange term is proportional to the inverse cube of the distance between two qubits with coupling strength $C_3$ which can also feature an angular dependency.
In contrast to the Ising model, excitations can hop from one site to another without the need for external driving (see Fig.\ref{fig1}(b)) but one can define a threshold distance $r_e = (C_3/\hbar \Omega)^{1/3}$ above which resonant driving dominates over the effect of spin-exchange. The spin nature of the excitations precludes the possibility of having two of them on the same site, resulting in an interactive version of a continuous quantum walk\,\cite{Childs_2003}. 

Remarkably, the blockade effect of the Ising or XY Hamiltonian can be used to encode the adjacency matrix of a graph. More precisely, one can assign qubits to vertices and their interactions to the presence or absence of edges in the graph.
This encoding is adapted to unit-disk (UD) graphs that can be defined with vertex positions and an edge between two vertices \textit{i.f.f.} the distance between them is smaller than a given threshold, taken as $r_b$ or $r_e$. In Figure\,\ref{fig1}(c), we show that the associated Ising interaction matrix closely reproduces the adjacency matrix of $\Graph$, a UD graph. However, because the interaction strength is proportional to $1/r^6$, next-neighbor interaction can still slightly contribute. We describe in Box~\ref{box:box1} how to mitigate such effects.

\begin{pasqbox}[label=box:box1]{Dealing with unsought terms in UD Graph embeddings}
The tail of the interactions can lead to spurious terms in the embedding of graphs, as illustrated by the lighter spots in right part of Fig.\,\ref{fig1}(c). The tunability of the qubits positions enables to partially alleviate their effects by optimizing atomic positions, as illustrated for example by the placement methods proposed in Ref.\,\cite{Atospatent,dalyac:tel-04265956}. However, when embedding onto a regular register, one needs to be aware of the ratio between nearest-neighbor distance $r_{nn}$ and next nearest-neighbor $r_{nnn}$ distance. 
For a theoretical clean embedding of a binary adjacency matrix where all edges have the same weight, the distances should be chosen such that $r_{nn}\gg r_b\gg r_{nnn}$. 
More practically, a good choice for the blockade approximation to work is $r_b\approx\sqrt{r_{nn}r_{nnn}}$.
For regular lattice or arbitrary configurations of atoms, having $r_{nnn}/r_{nn}\gtrapprox 1$ will translate into erroneous blockade approximation and thus smaller energy separation between blockaded and non-blockaded states.

More involved strategies could be used to entirely get rid of those unwanted terms.
One could combine dynamical decoupling/Floquet engineering techniques\,\cite{Vandersypen2004} together with Trotterization, in order to implement stroboscopically all the various Hamiltonian terms.
Another approach valid in the Ising encoding consists of taking advantage of the multi-level nature of atomic qubits.
One can thus use an additional non-interacting ground state to temporarily store the quantum state of a group of qubits, building on the idea of Ref.\,\cite{Bluvstein2022} where this approach was used for digital-analog computing.
Along the same line of thought, switching on and off desired interaction terms for selected durations can be achieved through Rydberg dressing\,\cite{Zeiher16,Hollerith2022}.
\end{pasqbox}

\newpage
The native interaction Hamiltonians (\ref{eq:ising_hamiltonian}) and (\ref{eq:xy_hamiltonian}) obtained in tweezer arrays platforms can be harnessed for tackling computational problems on graphs. It relies on the continuous evolution of the quantum state from an initial state $\ket{\psi_0}$ to 
\begin{equation}
   \ket{\psi(t)}=\mathcal{T}\left[\exp(-\frac{i}{\hbar}\int_{s=0}^{t}\mathcal{H}_\Graph(s)ds)\right]\ket{\psi_0}
   \label{eq:evolution}
\end{equation}
where $\mathcal{T}$ is the time-ordering operator and $\mathcal{H}_\Graph$ is the graph Hamiltonian, encompassing, in the Ising/XY model, properties related to nodes and edges of $\Graph$. On can thus encode the graph properties and use the evolution to produce graph-dependent quantum states, subsequently employed as resources for a variety of computing tasks.

\section{Combinatorial optimization on graphs}

\subsection{Solving the maximum independent set problem on unit-disk graphs}

The authors of Ref.\,\cite{Pichler2018} noticed that it was possible to leverage Rydberg blockade for exploring a well-known NP-hard combinatorial problem called the Maximum Independent Set problem (MIS) \,\cite{garey1978strong} on UD graphs.
Given a graph, the MIS problem consists in finding the largest subset of vertices which do not share an edge.
Numerous industrial challenges align with UD-MIS as comprehensively discussed in a recent review \cite{Wurtz22}. The ability to solve the MIS problem has broad implications, extending to various other graph-related problems, given their connection in graph theory.

The native use of interacting neutral atoms to explore the UD-MIS problem stems from the fact that the Rydberg blockade mechanism restricts the dynamics of interacting qubits to (coherent superpositions of) independent sets.
Setting $\Omega=0$ and carefully choosing $\delta>0$ in Eq.(\ref{eq:ising_hamiltonian}) enables to map the ground state of the Hamiltonian to the UD-MIS solution.
One can then use a system of driven interacting neutral atoms to (approximately) solve UD-MIS using a variational quantum adiabatic algorithm (VQAA)\,\cite{Kadowaki1998} or the quantum approximate optimization algorithm (QAOA)\,\cite{Farhi14,Zhou20}.
A few experimental implementations were performed\,\cite{Kim22,Ebadi2022} with the notable use of more than $280$ atoms in Ref.\,\cite{Ebadi2022}.
For long enough evolution times, the authors experimentally identified a quadratic quantum speedup with respect to classical simulated annealing on hard graphs.
Authors of Ref\,\cite{cain2023quantum} introduced an additional local Hamiltonian to the annealing scheme to achieve a quadratic speed-up over effective classical techniques.
Operations on atomic qubits are not limited to annealing schemes and recent works assess the use of quantum dynamics in excited regimes to solve combinatorial problems, using for example many-body localization\,\cite{wang2022many} or quantum critical dynamics\,\cite{king2023quantum}.
Specifically, quantum quenches with neutral atoms have been explored to circumvent super-exponential closing gaps in annealing schemes\,\cite{schiffer2023circumventing}.

Intriguingly, the locality of UD-MIS makes it relatively easy to approximate classically, and there even exists a so-called polynomial-time approximation scheme that guarantees a $1-\varepsilon$ approximation ratio in polynomial time\,\cite{hunt1998nc}, as already noticed in early publications\,\cite{Pichler2018,Serret20}.
Such algorithm guarantees to find a solution within a $1-\varepsilon$ factor of being optimal ($\varepsilon>0$) in a runtime polynomial in the problem size with exponent $1/\varepsilon$\,\cite{hunt1998nc}. Extrapolating from realistic coherence levels and current repetition rate of neutral atom devices, the authors of Ref.\,\cite{Serret20} estimated that a quantum algorithm exploiting this locality aspect could potentially surpass their classical benchmark with $8000$ atoms for a time budget of $2$ seconds, or with $1000 - 1200$ atoms for a time budget of $0.2$ seconds.
Those estimations call for a strong effort to increase register sizes of neutral atom devices. Some hardware elements related to the increase of atomic register sizes are described in detail in Box~\ref{box:box2}.

\subsection{Beyond the UD-MIS problem}

The existence of efficient classical approximation schemes for UD-MIS problems motivated efforts towards quantum procedures capable of tackling more complex groups of problems such as UD Maximum Weight Independent Set (UD-MWIS), where nodes are marked with weights. Weights represent an additional degree of complexity that can be tuned to make the instances harder to solve.
Employing site-dependent detuning, represented as $\delta_i$ in Eq.(\ref{eq:ising_hamiltonian}/\ref{eq:xy_hamiltonian}), facilitates encoding unique node weights of a graph into the Hamiltonian. This strategy offers a mechanism for encoding solutions to the UD-MWIS problem. However, the global driving $\Omega$ outlined in Eq.(\ref{eq:ising_hamiltonian}) proves inadequate in this context. Overcoming this limitation demands a more extensive set of operations, prompting the implementation of advanced sequences with local controls. The processes involved in achieving such addressing on the hardware are detailed in Box~\ref{box:box3}.

\begin{pasqbox}[label=box:box2]{Increasing atomic register sizes}

The techniques used to create the optical tweezers enable neutral atomic quantum processors to have inherent \textit{vertical scalability}, thus allowing the creation of larger graphs. This challenge involves fundamental hardware choices, notably on vacuum, optics and lasers. 

A good design of the vacuum chamber is critical for ensuring optimal optical access to the qubit register and low residual pressure, a requirement to extend the lifetime of atoms in optical tweezers. Typical experiments utilize compact vacuum chambers, often combined with a glass cell for optimal optical access. Achieving ultimate vacuum pressures involves cryogenic setups~\cite{pagano2018cryogenic,brandl2016cryogenic}. Current cryogenic neutral atom tweezers machines have lifetimes exceeding an hour, enabling the efficient assembly of matrices with hundreds of atoms~\cite{schymik2021single, schymik2022situ}. Current efforts aim to improve optics at low temperatures or integrate a cryopump into existing setups with extensive optical access.

In terms of optics, a crucial design decision revolves around choosing between \textit{ex-vacuo} and \textit{in-vacuo} systems, each presenting distinct advantages and disadvantages. Generating larger arrays requires wide-field microscope objectives with diffraction-limited performance over a large area and high numerical aperture. Customized optics, such as fixed masks (microlens arrays~\cite{pause2023supercharged}, metasurfaces~\cite{hsu2022single}, lithographic computer-generated holograms), offer solutions for improving the trap quality, at the expense of reconfigurability. 

The quantity of optical tweezers is directly proportional to laser power, but high powers can present difficulties for machine reliability and stability.  To improve the number of initially loaded traps and alleviate power constraints, adjustments such as starting from a colder atomic cloud and implementing shielding strategies have been explored~\cite{jenkins2022ytterbium,brown2019gray,aliyu2021d,shaw2023dark}. Additionally, optical lattices offer a power-efficient way of producing regularly spaced arrays, although optical tweezers are still needed to shuffle the individual atoms~\cite{young2020half,norcia2024iterative,tao2023high}. 

Exploring 3D graph production remains an intriguing frontier, with potential benefits from advanced vacuum systems, fast adaptive optics, and faster holographic imaging techniques~\cite{barredo2018synthetic,lee2016three,song2021quantum,schlosser2023scalable}.

A more demanding requirement is that the aforementioned advances should not compromise the overall performance of the system. Each step of the computational cycle, including assembly, state preparation, Hamiltonian evolution, and detection, has an associated infidelity that typically worsens as the number of atoms increases. Therefore, reducing SPAM errors should be a priority in future developments to ensure scalability without sacrificing performance.

\end{pasqbox}

\begin{figure}[ht!]
\center
\includegraphics[width=1.\textwidth]{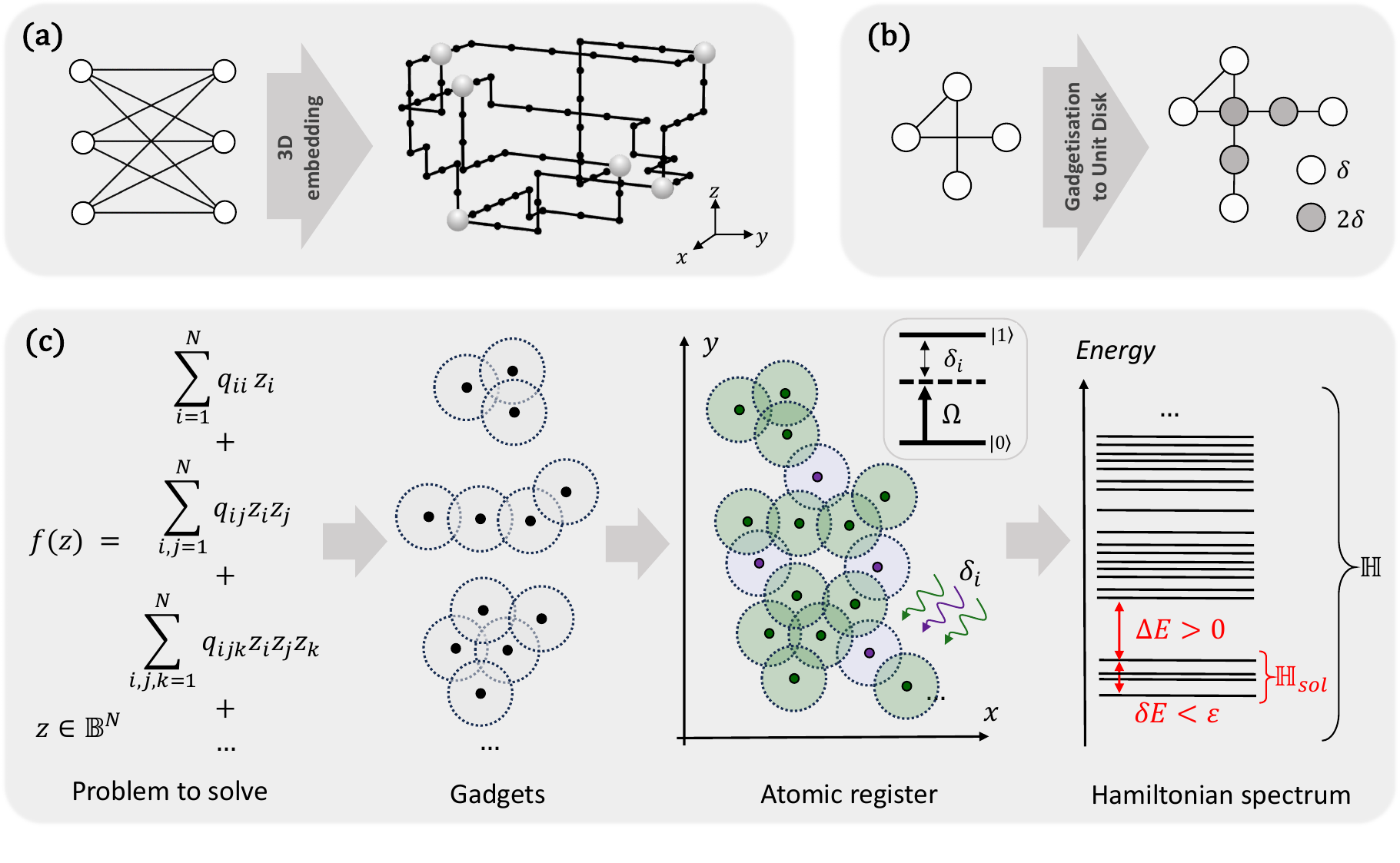}
\caption{Going beyond unit-disk using local addressing and ancillary atoms. \textbf{(a)} Embedding a non-planar graph in 3D using ancillary chains of atoms. \textbf{(b)} A 2D-gadget implementing a crossing edge using additional local detunings (grey nodes). \textbf{(c)} Framework of geometric programming: a general unconstrained binary optimization function $f$ is translated into a sum of terms that can each be encoded in an arrangement of $M$ atoms and local detunings $\{\delta_i, i \in 1, \dots, M \}$. Then, one can assemble these \textit{gadgets} together using pre-defined rules: this usually requires additional atoms (green). By construction, the Hamiltonian spans a Hilbert space where the low-energy bulk $\mathbb{H}_{\text{sol}}$ encodes the solutions to the QUBO in an $\varepsilon-$energy band. The excited states are gapped from the solutions by an energy difference $\Delta E > 0$.}
\label{fig:fig2}
\end{figure}

An alternative route involves the exploration of atomic graph embeddings that go beyond UD. It has indeed been observed that classical solvers exhibit a time-to-solution order of magnitude higher when applied to instances with larger connectivity or less local structure\,\cite{andrist2023hardness}. In particular, the incorporation of ancillary vertices, coupled with the ability to construct atomic arrays in 3D, facilitated the examination of the MIS problem on various platonic\,\cite{Byun22} and non-planar, high-degree graphs\,\cite{Kim22}. The key idea driving these developments is the utilization of a chain of auxiliary qubits to mediate strong interactions between distant qubits, allowing for the effective tackling of complex non-UD graphs. A subsequent demonstration in Ref.\,\cite{Dalyac2023} showcased a deterministic and polynomial approach capable of embedding any bounded degree graph. As an illustration, this method successfully addressed the MIS problem for the complete bipartite graph $K_{3,3}$ using a 3D arrangement of qubits shown in Fig.\,\ref{fig:fig2}(a), in conjunction with local detunings.

\begin{pasqbox}[label=box:box3]{Improving addressability in analog neutral atom QPUs}

To enhance the versatility of analog computing with neutral atoms, the focus is put on engineering local driving or detuning terms in the Hamiltonian~(\ref{eq:ising_hamiltonian}). Achieving this involves addressing individually the atoms with laser light, either resonantly for local driving ($\Omega$-term) or far-off resonance for local detuning ($\delta$-term) through the AC-Stark shift effect. Technically, this requires focusing a laser beam onto micron-sized spots overlapped with the atomic register. Controlling the intensity of each spot for maximal flexibility in Hamiltonian parameters poses a significant challenge due to the small scales involved and the need for fast temporal modulation at the nano-to-microsecond scale. Despite current limitations in achieving arbitrary qubit addressability, near-term implementations can combine a fast temporal modulation device with poor spatial resolution or flexibility, and a slower spatial modulation device.

Liquid-crystal based spatial light modulators (LCoS-SLM) offer a versatile method for the spatial shaping of addressing laser beams. However, these devices, while effective, face limitations in swiftly switching between configurations. Coupling LCoS-SLM with an electro-optical modulator (EOM) or an acousto-optical modulator (AOM) allows for temporal modulation, inducing an additional term in Hamiltonians~\eqref{eq:ising_hamiltonian}/\eqref{eq:xy_hamiltonian}
\begin{align}
\mathcal{H}_{add}(t)=A(t) \sum_j \epsilon_j \sigma_j^{\alpha}.
\label{eq:H_add}
\end{align}
The global temporal amplitude $A(t)$ is controlled by the amplitude modulator, while the LCoS-SLM spatially modulates the light intensity on all sites, providing site-dependent $\epsilon_j$. This technique is applicable to resonant or off-resonant light fields with $\alpha\in (x,y)$ or $\alpha=z$, driving addressed qubits or inducing a light shift, respectively. Notably, it has been used for tasks such as light shifting atoms for initial state preparation\,\cite{Scholl2022PRXQ} or during dynamics\,\cite{Chen2023} of an XY model and measurement of various spin components\,\cite{notarnicola2021randomized}.

Alternative approaches for addressability involve acousto-optical deflectors (AODs), either alone or combined with LCoS-SLM, allowing neutral atom addressing~\cite{barnes2022,Bluvstein2023}. Shuttling atoms with AODs to a dedicated zone for addressing with a fixed laser beam is another strategy~\cite{Bluvstein2022}. Higher refreshing rates for spatial modulation are achieved using Micro-electromechanical systems (MEMs) or Digital micromirror devices for local Raman transitions or Stark-shift addressing on neutral atoms or ions~\cite{knoernschild2010,zhang2023scaled}. Multiplexing approaches, utilizing individual modulators for each channel, can be applied in free space with a multichannel AOM or using photonic integrated circuits (PICs)\,\cite{wright2019,menssen2022}. The latter case incorporates a steering SLM to match the pattern out of the PIC to the atomic register\,\cite{christen2022}. Despite the challenges, these techniques provide diverse options for achieving addressability in neutral atom systems.

\end{pasqbox}

Advancements in local controls and ancillary chains of atoms have spurred the utilization of the native UD-MIS solving capabilities of neutral atoms to address optimization problems in higher complexity classes. Solutions to the original problem are mapped to the lower energy states of the Rydberg Hamiltonian, a process referred to as \emph{geometric programming} on Rydberg platforms~\cite{Stastny2023completeness}. This involves finding a geometric arrangement of atoms and a set of laser controls to achieve the desired hierarchy of eigenstates. Various methods of geometric programming have been recently proposed, relying on \emph{gadgets}—small atomic ensembles encoding parts of an optimization problem. Assembling these gadgets using defined gluing rules constructs a faithful UD-MIS problem. For instance, Ref.\,\cite{Nguyen22} introduces copy and crossing gadgets that facilitate all-to-all connectivity on a crossing lattice, as illustrated in Fig.\,\ref{fig:fig2}(b). This enables the encoding of any unweighted Quadratic Unconstrained Binary Optimization (QUBO) problem into a UD-MWIS problem, using at most $4N^2$ atoms and global detuning masks. Gadgets described in Ref.\,\cite{lanthaler2023rydberg} specifically address the implementation of 3- and 4-body parity constraints in the parity encoding of connected higher-order binary optimization (HOBO) problems, facilitating the encoding of any HOBO as a UD-MWIS.

Additional gadgetization techniques, such as the QUBO-centric approach proposed in Ref.\,\cite{byun2023rydbergatom}, offer robust solutions without requiring local fields. Triangular gadgets and anti-copy wires, introduced in Ref.\,\cite{Jeong2023}, encode 3-Satisfiability (3-SAT) problems as UD-MIS problems, demonstrating experimental success on small systems. This toolbox has been experimentally applied to factorize small semi-prime numbers in~\cite{park2024rydbergatom}. Geometric programming techniques have also been developed for quantum simulation, expanding the range of algorithmic problems solvable by neutral atoms, as seen in~\cite{Stastny2023completeness} or~\cite{zeng2024quantum}. The geometric programming formalism is underpinned by the functional completeness of PXP models, as demonstrated in~\cite{Stastny2023completeness}. The hard-blockade approximation, enabled by an embedding respecting $r_{nn}\ll r_b\ll r_{nnn}$, allows encoding solutions of various algorithmic problems into the bottom end of a Rydberg Hamiltonian spectrum with a finite gap and perfect degeneracy, as depicted on the right of Fig.\,\ref{fig:fig2}(c). These techniques significantly broaden the spectrum of algorithmic problems solvable by neutral atoms, albeit at the cost of introducing auxiliary atoms to propagate Rydberg blockade and employing local controls to fix different maximum independent configurations at the required energy levels on the Hamiltonian spectrum.

\section{Graph Machine Learning}

As seen previously, it is possible to replicate to some extent the adjacency matrix of a graph with the interaction matrix of Rydberg atoms. Graphs are not only suited to represent combinatorial problems, but they are also a natural representation of data in many domains of sciences. They naturally describe relationships in social networks\,\cite{Freeman2000VisualizingSN}, characterize interactions of proteins and genes\,\cite{Theocharidis2009NetworkVA} and can also represent the structure of sentences in linguistics\,\cite{sole2001small}. Graph machine learning (GML) aims at extracting valuable
patterns from complex graph data. Many impactful applications arise from efficient GML methods, such as predicting potential edges in recommendation systems\,\cite{Schafer01}, detecting frauds in communication networks\,\cite{Pourhabibi20}, or for protein function prediction\,\cite{Muzio20}. An important research topic is to therefore find efficient graph embedding techniques\,\cite{goyal2018graph}, which refer to finding a representation of a graph or of its individual nodes in a vector space. Those embeddings can then be used for prediction tasks at the node-level, such as node classification\,\cite{Bhagat2011}, link prediction\,\cite{LibenNowell03} or at the graph-level to distinguish graphs of different nature. 
Using the exponentially large Hilbert space accessible to a quantum computer in order to generate graph embeddings is an appealing idea, with many proposals and theoretical studies over the past few years\,\cite{Schuld19,Havlek19,schuld2020measuring,kishi21}.

\subsection{Quantum feature maps}
\begin{figure}
    \centering
    \includegraphics[width=\textwidth]{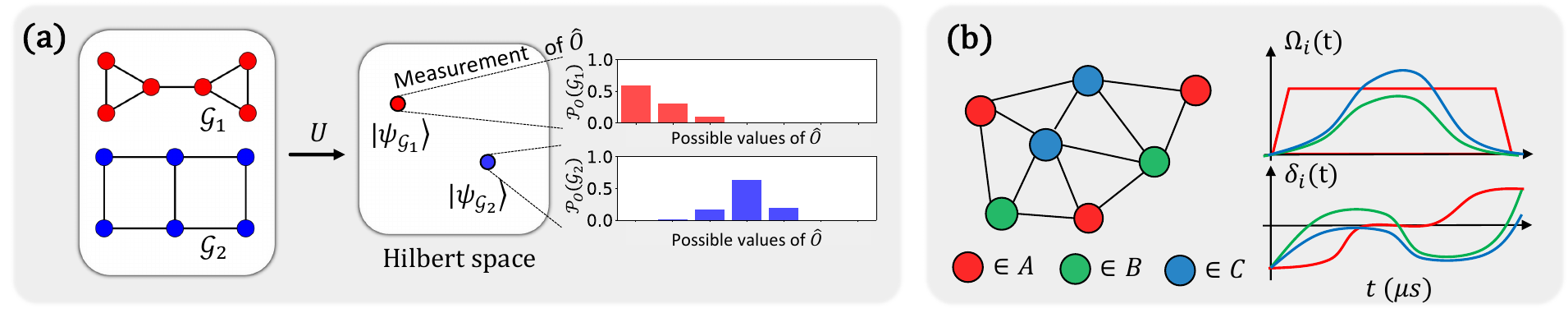}
    \caption{ \textbf{(a)} A parameterized quantum unitary $U$ is applied to atomic registers arranged under the form of UD graphs $\Graph_{1/2}$. Experimentally measured observable distributions $\mathcal{P}_O$ are then used as modified inputs for learning tasks. \textbf{(b)} Nodes with distinct labels $\{A,B,C\}$ in a graph can be addressed with specific local controls, an especially useful feature for  GML algorithms.}
    \label{fig:GML}
\end{figure}

Recent works have proposed the utilization of neutral atoms for implementing quantum graph machine learning approaches\,\cite{Henry2021}, building upon theoretical studies\,\cite{Cong2019,Verdon2019} and a proposal focused on a photonic platform\,\cite{schuld2020measuring}. These procedures involve encoding graph data in a quantum state, realizing a {\it quantum feature map}\,\cite{Schuld19, Havlek19} as illustrated in Fig.\ref{fig:GML}(a).

The concept introduced in\,\cite{Henry2021} involves leveraging the quantum dynamics of a collection of atoms organized in a 2D graph-like structure, followed by the measurement of an observable on these atoms. Through these measurements, it becomes possible to extract characteristic features for each graph. In particular, one can derive probability distributions that serve as fingerprints of the graph. 
One of the principal advantages of encoding the topology of the graph directly into an atomic register is its native equivariance, a fundamental requirement for any model aiming to incorporate an inductive bias enabling generalization to unobserved data\,\cite{geometric_DL}. Recent attempts have explored this idea in quantum machine learning with notable success\,\cite{Alamos2, mernyei2022equivariant, skolik2023equivariant}.

Careful choices of the observables to be measured allow probing various aspects of the graph while preserving the desired symmetries through measurement histograms. Analyzing the characteristics of these distributions enables the differentiation of distinct types of graphs. Notably, discriminative models can be trained by tuning the driving parameters. A first proof-of-concept for an experimental demonstration was reported in Ref.\,\cite{Albrecht2023}, in which the authors used quantum graph embeddings for a toxicity screening procedure on a standard biochemistry dataset comprising $286$ graphs of sizes ranging from $2$ to $32$ nodes. Despite using the processor with global addressing only, this procedure achieved classification accuracies on par with state-of-the-art classical kernels. More importantly, the authors of Ref.\,\cite{Albrecht2023} demonstrated that the geometry induced by the quantum procedure was challenging to replicate using classical means, employing a geometric difference tool introduced in Ref.\,\cite{Huang2021}, thereby opening the way to potential representation advantages.

\subsection{Limitations and prospects for quantum GML}

One notable limitation in previous experiments stems from the reliance on global driving. In graph theory terms, this implies that only the adjacency matrix of the graph has been considered. Real-life graphs, however, often encompass vertex and edge information, which significantly contribute to the graph's functionality. For example, when examining molecules, the types of atoms and chemical bonds influence a molecule's reactivity, and incorporating this information is essential in classical graph machine learning methods\,\cite{Gilmer2017}. To address this need, nodes with distinct labels in a graph can be managed with specific local controls, as illustrated in Fig.\,\ref{fig:GML}(b).

Another substantial limiting factor is the relatively low repetition rate of neutral atom devices and other Quantum Processing Units (QPUs). Contemporary machine learning approaches routinely handle datasets comprising $10^5-10^6$ graphs, all of which would need to be encoded on the device at least once. The sluggish data encoding into neutral atom QPUs, resulting from their low clock rates, is crucially needing improvements. Hardware elements related to enhancing this repetition rate are detailed in Box~\ref{box:box4}.

A final noteworthy limitation arises from the relatively small maximum size of graphs that can be encoded using this approach. While some problems or datasets involve graphs of modest size (around $100$ nodes), many others target much larger graphs. A promising strategy to contend with the finite number of qubits available in today's QPUs involves partitioning complex graph problems into smaller, manageable components. This allows for simplifying a graph while retaining its fundamental structure. Machine learning approaches such as graph generative models can be employed for this purpose, tailoring subgraphs with specific edge types that are then aggregated to reconstruct the initial graph. This methodology can be extended to heterogeneous graphs—graphs encompassing varied node and edge types—by fragmenting them into a composite of homogeneous graphs.

These experimental developments stand to benefit from recent advances in geometric quantum machine learning\,\cite{Larocca2022,Skolik2022}, with the latter exploring applications to weighted graphs. Furthermore, in-depth theoretical studies on equivariant and geometric aspects of quantum machine learning have been presented recently\,\cite{Alamos1,Alamos2,Alamos3}.

\begin{pasqbox}[label=box:box4]{Increasing the clock rate}
A major limitation of neutral atom technology naturally arises when summing the duration of all the required steps of the processing cycle. While the actual quantum dynamics happens at typical MHz scale, the necessity to load atoms in the traps ($\sim 100$ms), arrange them into a specific spatial graph configuration($\sim 100$ms) and image them several times ($\sim 50$ms) at each shot, reduces the repetition rate of the useful computation to few Hz. In addition, the effective repetition rate is even scaled down by the finite probability of assembling a defect-free quantum register at each cycle. As moving an atom from one trap to another has a finite efficiency $\eta$ due to losses, the probability of reaching a perfect arrangement of $N$ atoms scales as $\eta^N$. Thus, at $\eta=99\%$, arranging a $100$-node graph would take $1/\eta^{N}\approx3$ times more cycles than with loss-free moves ($\eta=1$).

Boosting the repetition rate to start closing this frequency gap with others technologies thus carries substantial significance and requires technical improvements at each step. The loading stage can be fastened by continuously loading the target register using an atom reservoir \cite{gyger2024continuous,norcia2024iterative} or improving the filling efficiency from $50\%$ to $\sim90\%$ using gray molasses \cite{PhysRevX.9.011057}.
The scaling of the duration required by an atom-by-atom assembler is another bottleneck which can be addressed by either parallel transport of atoms \cite{Bluvstein2022} or optimized arrangement algorithms \cite{Cimring_2023}. 
Fast imaging can be obtained using resonant imaging of free atoms with high numerical aperture optics \cite{Bergschneider_2018} or replacing electron-multiplying charge-coupled devices (EMCCD) used to collect fluorescent photons by cryogenic avalanche detectors \cite{Buzulutskov_2012}. One can ultimately achieve fast high-fidelity non-destructive measurement by using cavity-assisted detection such as in \cite{Deist_2022}. Overall, the combination of all those techniques should raise the effective repetition rate to the 100 Hz or kHz range.

\end{pasqbox}

\subsection{Quantum Extremal Learning}
\label{sec:QEL}

So-far in this section, we focused on solving optimization tasks where the mathematical cost function had a well-defined cost associated with each input, such as QUBOs.
In this setting, the real-world problem has a known interpretation or conversion into such model and one can derive the numerical values of each of the terms, such as the edge and node weights in a graph optimization task. The only challenge lies in finding which input would minimize the output of this cost function. 

However, a known mathematical model describing the observed phenomenon is not always provided.
The system may simply be too complex to model or its inner workings may remain hidden for various reasons.
Instead, only a dataset of input-output relationships, or "measurements" on the system, could be given. Relying solely on this dataset, the primary objective remains, as before, to identify an extremal value of this relationship taken as figure-of-merit (FOM).
One would not just be interested in the extremum among the measured data points but would first infer and generalize the relationship, to be able to find a even better extremum and its corresponding input.

This scheme called "extremal learning" involves in the first step to learn a surrogate function based on a given dataset, and at a second step to optimize this function to find its extremal value. 
Previous methods \cite{kitai2020, Matsumori2022} relied on classically fitting QUBOs to datasets first, using for instance Factorization Machines, and then optimizing with a quantum QUBO solver.
The authors of Ref.\cite{varsamopoulos2022quantum} uses a quantum machine learning model as a surrogate before applying the quantum optimization procedure, greatly extending the class of models that can be studied and optimized. One such class of models apply to graph-structured datasets where inputs are graphs and outputs are scalar values. An example of an industrially-relevant dataset was presented in \cite{varsamopoulos2022quantum}, where molecular structures, with atoms being nodes and chemicals bonds being edges, and FOMs were correlated. 
As molecular graph structures can be embedded in Euclidean geometrical space, one can fit a QUBO or a more general quantum model to the dataset directly. This method exemplifies how classical ML and quantum optimization techniques can be combined in a hybrid fashion for graph-structured dataset analysis. In the following section, we explore various examples of such hybrid schemes for both combinatorial optimization and machine learning tasks.

\section{Hybrid quantum-classical algorithms}

State-of-the-art classical solvers employ extensive pre-processing and heuristics rather than brute force methods to efficiently tackle large-scale problems. For example, branch-and-bound (B$\&$B) methods\,\cite{morrison2016branch} divide the input into smaller problems and employ bounds to eliminate sub-problems that cannot lead to an optimal solution. Kernelization, on the other hand, reduces the size of the input while preserving the solution and helps to distill the essential information of the problem into a more manageable size\,\cite{fomin2019kernelization}. Methods such as B$\&$B and kernelization therefore offer an elegant framework to limited quantum processors. They enable quantum processors to handle limited-size components of a problem while delegating the pre-processing and the aggregation of sub-solutions to classical processors. Implementing "non-native" hybrid schemes has recently been proposed on neutral-atom quantum computers\,\cite{wurtz2024solving}.

\begin{figure}
    \centering
    \includegraphics[width=\textwidth]{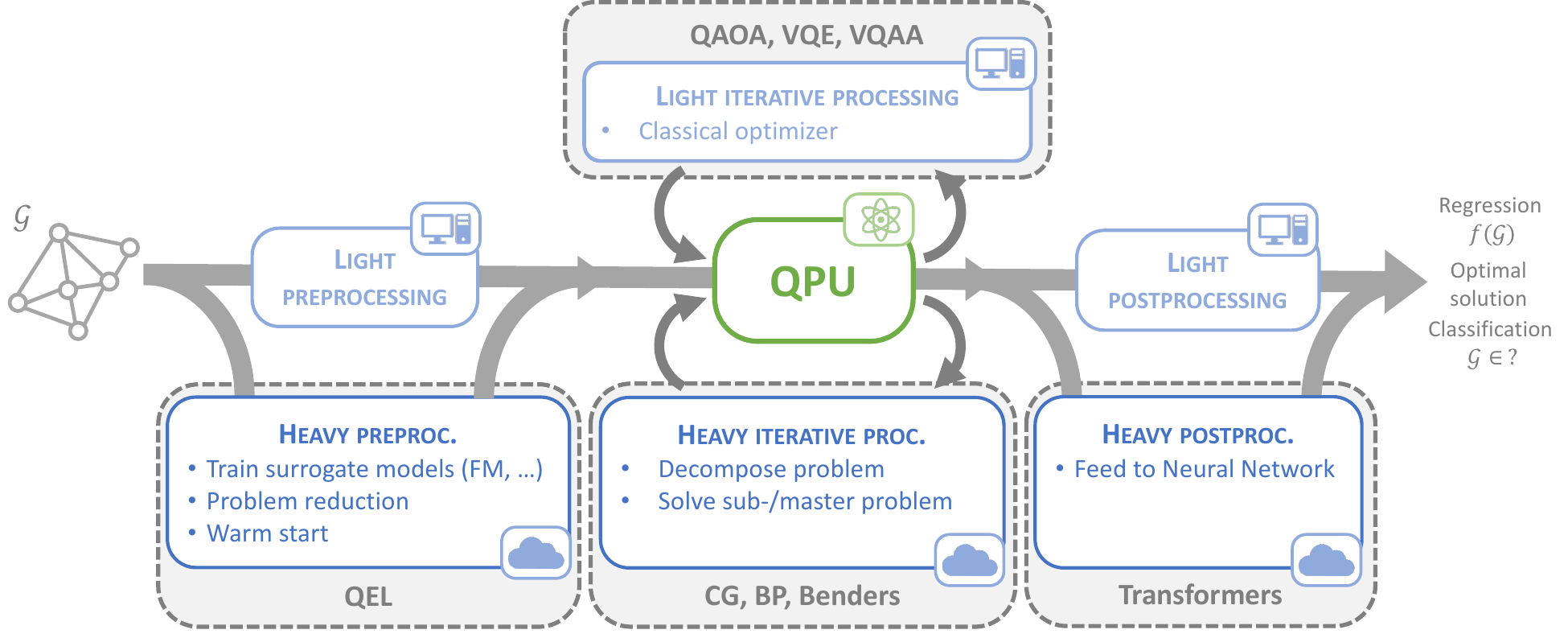}
    \caption{Schematics for possible hybrid algorithms.
    These algorithms require heavy classical computation, either at preprocessing (left) as detailed in \ref{sec:QEL}, in iterative processes (center) as detailed in \ref{sec:hybridQGO} or at post-processing (right) as detailed in \ref{sec:hybridGML}.
    }
    \label{fig:hybrid}
\end{figure}

\subsection{Hybrid optimization approaches}
\label{sec:hybridQGO}

In the realm of classical optimization problems, the quantum optimization framework can be a standalone solution or a module integrated in a hybrid workflow. Accordingly, we categorize hybrid quantum-classical algorithms into two main groups.

In the first category, solving the combinatorial problem already embedded into an atomic register amounts to an optimal control task. The algorithm involved may rely on a set of parameters, such as pulse shape or circuit layer, which when optimized gives an efficient protocol to drive the quantum system to the correct solution state. 
In this context, as shown on top of Fig.\ref{fig:hybrid}, a light classical routine performs an optimization process over the parameters of the controls, in closed loop with the QPU. A substantial portion of quantum algorithms falls within this class, including variational algorithms\,\cite{cerezo2021variational} and quantum neural networks\,\cite{schuld2014quest}.

The second category of hybrid classical-quantum methods involves decomposing the optimization problem at hand into two or more sub-problems, each linked to specific decision-making processes of the original problem. While certain sub-problems are tackled by a quantum algorithm such as those belonging to the first category, others are resolved using classical resources. The overall solution to the original problem is then a composite of the solutions to each sub-problem. To differentiate from the previous category, one can examine the algorithm after parameter setup: if some decision-making steps are still carried out by a classical algorithm, it belongs to the second category, known as hybrid algorithms. The authors of Ref.\cite{PhysRevResearch.5.043117} implements such an hybrid algorithm to construct a quantum-enhanced classifier for the prediction of credit rating downgrades. The QUBO-solving block of the QBoost-based algorithm \cite{neven} utilizes a neutral atom QPU with either QAOA-like method or quantum sampling.  Over the years, several advanced methods have been proposed, each with significant potential and notable limitations in terms of universality. 

A widely applied classical method is the \textit{Column Generation algorithm}\,\cite{desrosiers1984routing}, which aims to constrain the solution space to a few viable options. It iteratively adds new options during the process only if they can enhance the current best solution for the problem. However, identifying such solutions involves addressing several smaller, yet challenging, combinatorial sub-problems. Therefore, designing an efficient approach to solve these sub-problems, commonly known as pricing sub-problems, is crucial to ensuring high-quality solutions for a given problem.

Researchers in\,\cite{da2023quantum} demonstrated that addressing the pricing sub-problems with a quantum algorithm could not only enhance the quality of the final solutions but also reduce the number of iterations needed to find them. Utilizing a neutral atom-based quantum computer as a sampler, one can design a quantum algorithm to identify not only one but several viable options to be added to the original problem. The authors presented a proof-of-concept for solving the well-known Vertex Coloring problem, which has immediate real-world applications such as resource allocation, scheduling, and planning-related problems\,\cite{malaguti2011exact}. 

Benders' Decomposition (BD) is another well known method in mathematical optimization for solving large-scale problems, especially mixed-integer linear programs (MILPs) with continuous and integer variables. It breaks down the problem into a master problem and subproblems. The master problem deals with only a subset of decision variables, while the subproblems handle the rest, and usually becomes easier to solve. Solutions from subproblems generate additional constraints, called Benders' cuts, enhancing by addition the master problem's solution. While the subproblems are in general manageable on classical computers, the master problem, containing discrete variables, constitutes the computational bottleneck. Authors in \cite{naghmouchi2024mixed} propose a framework combining classical BD with neutral atom-based computation where they automate the conversion of the master problem into a QUBO formulation.

\subsection{Hybrid GML approaches}
\label{sec:hybridGML}

Similar to the quantum optimization approach, we anticipate significant advancements in both performance and insights through the hybridization of QPUs with classical computers in GML. Utilizing the unique expressiveness of quantum feature maps in conjunction with state-of-the-art graph machine learning techniques can lead to fruitful developments\,\cite{QGNN_hybrid_1,QGNN_hybrid_2,thabet2022extending}. These concepts have recently been explored in\,\cite{thabet2023enhancing}, where the authors employ quantum correlation matrices as positional encoding for a transformer model. The method demonstrates promising results on both synthetic and real benchmark datasets.

In\,\cite{thabet2023enhancing, Gamble_2010}, empirical evidence is provided to show that the use of quantum features, particularly in the case of $k=2$ particle quantum walks ($k$-QWs), exhibits superior expressiveness. When tested on arbitrarily chosen families of non-isomorphic strongly regular graphs, the $k$-QWs can distinguish them, whereas classical walk kernels and $k$-Weisfeiler Leman ($k$-WL) tests with $k<3$ (at least) fail. It's worth noting that classical $k$-WL involves comparisons between $k$-tuples of nodes, rendering its complexity at least $\mathcal{O}(N^k)$, where $N$ is the size of the graph. In contrast, for $k$-QWs, the algorithm's complexity is determined by the number of shots needed to measure and reconstruct the distribution of the desired observable. The number of shots is $\mathcal{O}(\frac{1}{\epsilon^2})$ for a precision up to $\epsilon$\,\cite{huang2020predicting}. This potential advantage arises if the expressiveness of $k$-QWs is provably comparable to $k$-WL, as observed empirically for $k$ equal to 1 \& 2 but not proven in the general case.

This approach is not limited to transformer architectures alone. Quantum feature maps can be integrated into a generic encoder, which undergoes classical post-processing in any machine learning framework, as long as it ensures superior expressiveness for the specific task. In the realm of graph machine learning, this translates to the distinguishability of graphs by the WL test and its variants\,\cite{Morris_Ritzert}. This allows us to harness the potential of quantum computers in the current Noisy Intermediate-Scale Quantum (NISQ) era while processing data at scale.

\section{Conclusion and perspectives}

The rapid advancements in neutral atom technology present a thrilling opportunity for addressing graph-related challenges. By harnessing the current and prospective capabilities of hardware, we have demonstrated the feasibility of embedding a wide range of graph-related problems into the Hamiltonian of interacting neutral atoms. Additionally, we have put forth suggestions for hardware enhancements, foreseeing an increase in atomic register sizes, addressability, and clock rates of neutral atom devices in the near future. The integration of these anticipated upgrades in classical-quantum approaches not only pushes the boundaries of combinatorial optimization and graph machine learning but also holds promising prospects for practical real-world applications.

Neutral atom technology is not merely narrowing the gap with other platforms but is also displaying substantial potential for further improvement. The rapidly expanding network of neutral atom researchers, which has grown from a mere three academic groups two decades ago to nearly two hundred today, is actively exploring innovative approaches with exciting features. Noteworthy developments include emerging platforms utilizing atom shuttling\,\cite{Bluvstein2023}, non-destructive detection\,\cite{Deist_2022}, alkaline-earth atoms\,\cite{madjarov2020high,jenkins2022ytterbium,Ma2023}, bi-species arrays of rubidium and cesium atoms\,\cite{singh2022dual}, circular Rydberg levels\,\cite{ravon2023array}, or a combination of tweezers technology and cryogenic environments\,\cite{schymik2022situ}. These novel methods have the potential to provide all-to-all connectivity and extended coherence times achieved through faster entangling protocols or by preventing state contamination caused by black-body radiation. In essence, the flourishing neutral atom community is paving the way for a multitude of exciting scientific discoveries on the horizon.
\printbibliography
\end{document}